# Compact, all-PM fiber integrated and alignment-free ultrafast Yb:fiber NALM laser with sub-femtosecond timing jitter

Yuxuan Ma, Sarper H. Salman, Chen Li, Christoph Mahnke, Yi Hua, Stefan Droste, Jakob Fellinger, Aline S. Mayer, Oliver H. Heckl, Christoph M. Heyl, and Ingmar Hartl

*Abstract*—We report a simple and compact design of a dispersion compensated mode-locked Yb:fiber oscillator based on a nonlinear amplifying loop mirror (NALM). The fully polarization maintaining (PM) fiber integrated laser features a chirped fiber Bragg grating (CFBG) for dispersion compensation and a fiber integrated compact non-reciprocal phase bias device, which is alignment-free. The main design parameters were determined by numerically simulating the pulse evolution in the oscillator and by analyzing their impact on the laser performance. Experimentally, we achieved an 88 fs compressed pulse duration with sub-fs timing jitter at 54 MHz repetition rate and 51 mW of output power with $5.5 \times 10^{-5}$ [20 Hz, 1 MHz] integrated relative intensity noise (RIN). Furthermore, we demonstrate tight phase-locking of the laser's carrier-envelope offset frequency ($f_{ceo}$) to a stable radio frequency (RF) reference and of one frequency comb tooth to a stable optical reference at 291 THz.

## I. INTRODUCTION

Mode-locked ultrafast laser oscillators have numerous applications in industry and science. Ultrafast oscillators in the Yb-gain spectral region around 1 µm routinely drive high-power Yb-amplifiers [1], [2], powering the growing field of ultrafast applications with simultaneous need for high repetition rate and high pulse energy. Examples are material processing [3], superconducting free-electron lasers (FELs) [4], strong-field and attosecond physics [5], Terahertz (THz) generation [6], and optical frequency comb (OFC) spectroscopy ranging from the extreme-ultraviolet (XUV) [7] to mid-infrared (MIR) [8]. In recent years, ultrafast fiber oscillators based on a NALM [9] as a saturable absorber (SA) have gained attention due to their high stability and low-noise properties. Those devices can be highly reliable due to the use of PM fibers and the fact that no semiconductor SAs are required, an element which often shows degradation over time [10]. Initially, NALM oscillators were constructed with Er-doped gain fibers and two fiber loops in a figure-8 configuration [11]–[13]. Later, figure-8 oscillators in the Yb-gain region were reported [14], [15]. In this configuration it is difficult to directly obtain sub-100 fs pulse duration and high repetition rates ($f_{rep}$). To overcome these drawbacks a novel laser configuration comprising a single loop and a linear arm was proposed [16]–[20]. This configuration requires a non-reciprocal phase bias for mode-locking [21]. For this laser design, extensive research was conducted to investigate the noise properties [22], to increase the $f_{rep}$ [23] and the output power [24], to stabilize the $f_{ceo}$ [16], [25], [26], to make the laser structure more compact [18], and to utilize it in many applications, including space-borne science [27].

While ultrafast laser systems delivering the highest powers today are based on various Yb-doped host materials around 1 µm, NALM mode-locked Yb:fiber seed oscillators needed for those systems are not as mature as their Er:fiber-based counterparts in terms of reliability and robustness. One major limiting factor in previously demonstrated Yb-NALM oscillators is the use of free-space optics for dispersion management and non-reciprocal phase shift. Fiber dispersion-management methods cannot simply be transferred from the 1.5 µm Er-gain region to 1 µm due to the lack of anomalous dispersive single-mode step-index fibers at 1 µm. A compact Yb:fiber NALM laser design [18] without dispersion compensation was proposed by T. Jiang *et al.*, however the trade-off is high phase noise, a narrow spectral bandwidth (3.1 nm) and long pulse durations (538 fs).

Here, we demonstrate a rugged Yb:fiber NALM oscillator design without any tunable or alignable parts. Since the pump power is the only degree of freedom, our design provides dramatically enhanced reliability, robustness and long-term stability compared to previous demonstrations. The oscillator generates sub-100 fs pulses at low intensity noise and timing jitter and allows $f_{ceo}$ stabilization for OFC applications.

The paper is organized as follows: In section II we present simulation results to illustrate the influence of various design parameters on the performance of the oscillator. In section III we present the experimental realization of the oscillator design and the achieved performance. Sections IV and V describe the noise characterization and frequency comb operation, respectively.

## II. NUMERICAL SIMULATIONS

Figure 1 shows our oscillator design which generally consists of a NALM and a linear arm. The NALM comprises a 2 × 2 type fiber coupler, an asymmetrically placed Yb-doped gain fiber pumped by a 976 nm fiber coupled laser diode (LD) and a non-reciprocal phase shifter. In the linear arm, a CFBG acts as the cavity end mirror and the dispersion compensation element simultaneously. All components are connected via single-mode passive fibers (PM980). We simulated this oscillator design using a commercial software package [28]. The simulation was started by placing a random and weak initial noise field numerically inside the cavity. This field was then propagated through each individual component in the cavity one by one. We repeated the oscillator round-trip iterations until the relative change of pulse duration between adjacent round-trips was below $10^{-7}$. At this level, we assumed convergence into a stable mode-locking regime, which typically happened after several hundreds of iterations.

Table 1 lists the parameter range we used in our simulations. The oscillator repetition rate $f_{rep}$ is kept fixed at 54.17 MHz in all simulations, due to requirements in one of our target applications [29]. We also kept the net intracavity dispersion fixed at $-0.01$ ps$^2$ for short pulse duration and low-noise operation. Finally, we kept the gain fiber length fixed at 0.6 m. We had optimized this length experimentally in similar Yb:fiber oscillators prior to the simulations.

Our simulations include all major effects such as fiber dispersion, nonlinear refractive index, transverse power distribution in the fiber, dynamic gain, and optical interference. In the remainder of this section we will discuss the influence of several key design parameters on the laser performance. For each parameter under study (see table 1 for abbreviations) we performed dozens of simulations where we varied the parameter while keeping all other parameters fixed.

TABLE 1
VARIABLES IN THE NUMERICAL SIMULATIONS

| Symbol | Description | Typical value (used in experiment) |
|---|---|---|
| $r$ | Power splitting ratio of fiber coupler | 10% ~ 90% (50%) |
| $R_p$ | Peak reflectance of the CFBG | 20% (fixed) |
| $BW_{CFBG}$ | FWHM bandwidth of the CFBG | 5 nm ~ 40 nm (20 nm) |
| $GDD_{net}$ | Intra-cavity net dispersion | $-0.01$ ps$^2$ (fixed) |
| $\Phi_0$ | Non-reciprocal phase bias | 0 ~ 2$\pi$ ($\pi$/2) |
| $L_1$ | Passive fiber length (the short part in NALM) | 5 cm (fixed) |
| $L_2$ | Passive fiber length (the long part in NALM) | 0.45 m ~ 1.95 m (1.35m) |
| $L_a$ | Active fiber (Yb 401-PM) length | 0.6 m (fixed) |
| $L_3$ | Passive fiber length (the linear arm) | Accordingly (~80 cm) |
| $f_{rep}$ | Repetition rate | 54.17 MHz (fixed) |
| $P_p$ | 976 nm pump power | 0 ~ 400 mW (150 mW) |

### A. Non-reciprocal phase bias

The non-reciprocal phase bias [21] $\Phi_0$ is one of the most essential elements for starting and maintaining mode-locking of a NALM oscillator. For stable mode-locking, SA action is required, i.e. the cavity loss needs to decrease with increasing pulse peak power. However, for many choices of $\Phi_0$ mode-locking is not obtainable either because the cavity loss increases with increasing peak power or because the loop mirror reflectance is too low for sustaining cavity oscillations at the initially low peak powers when starting up from noise.

We simulated $\Phi_0$ in the range between 0 and 2$\pi$ with a step of 0.1$\pi$ (assuming the following fixed parameters: $r$ = 50 %, $BW_{CFBG}$ = 20 nm, $L_2$ = 1.35 m, and $P_p$ = 80 mW). The general trends are shown in Fig. 2: In the range from 0.1$\pi$ to 0.7$\pi$, with increasing $\Phi_0$, the oscillator is less likely to self-start, but double pulsing is better suppressed. Also, the power at the output port decreases, while the power at the rejection port increases and the output spectral bandwidth decreases. For other $\Phi_0$ values, the simulation couldn't converge to a steady state, therefore we label these cases as "no mode-locking". Due to the tendency to double-pulse at low $\Phi_0$, the useable range of $\Phi_0$ for our design is between 0.4$\pi$ and 0.7$\pi$. For the best compromise between output power and spectral bandwidth, we choose a value of $0.5\pi$ for the phase bias $\Phi_0$ in the experimental realization of the laser.

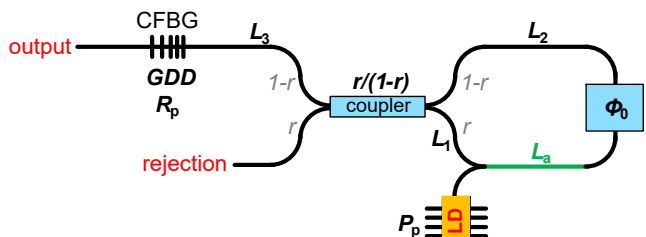

Fig. 1. Schematic of the laser model for numerical simulation. See table 1 for the definitions of the parameters.

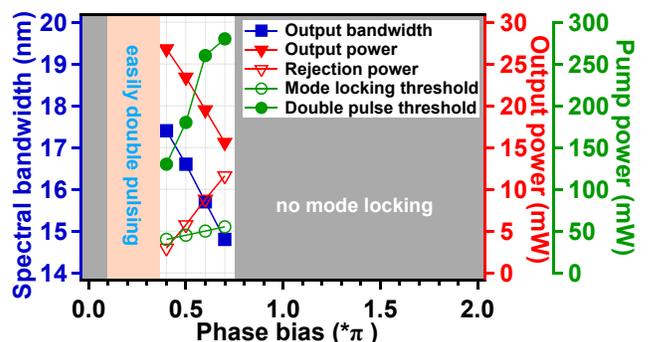

Fig. 2. Simulation results: influence of non-reciprocal phase bias $\Phi_0$ on laser performance. Fixed parameters: $r$ = 50 %, $BW_{CFBG}$ = 20 nm, $L_2$ = 1.35 m, and $P_p$ = 80 mW.

*B. Coupler power splitting ratio*

The coupler splitting ratio $r$ mainly influences the modulation depth of the artificial SA, i.e. the NALM. At increasing deviation of $r$ from 50 %, the modulation depth is decreasing [19]. On the other hand, for higher $r$, more power will propagate in the counter-clockwise direction, resulting in a larger nonlinear phase difference between the counter-clockwise direction and clockwise direction, as shown in Fig. 3. Note that if $r \neq 50$ %, it matters which of the coupler ports is spliced closer to the gain fiber, since this will determine whether the pulses with higher energy really experience a higher nonlinear phase shift or vice versa.

Generally, when $r$ is close to 50 %, the output power and spectral bandwidth is relatively insensitive to $r$. However, these two parameters deteriorate when $r$ is above 60 %. Finally, mode-locking is lost for $r > 85$ %. We chose to use a 50 % coupler in the experiment for the highest output power, broadest spectral bandwidth and larger modulation depth. The latter is beneficial to suppress continuous wave lasing breakthrough after mode-locking is achieved.

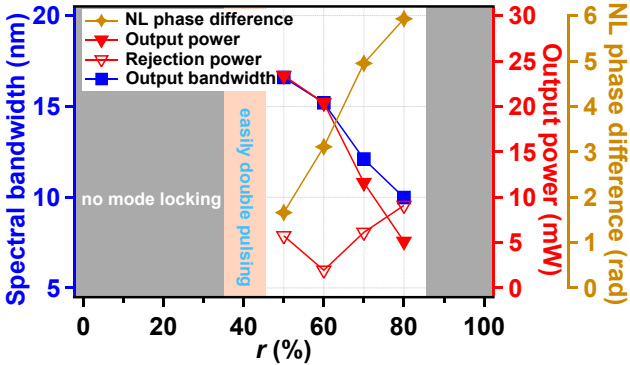

Fig. 3. Simulation results: influence of coupler splitting ratio on laser performance. Fixed parameters: $\Phi_0 = 0.5\pi$, $BW_{CFBG} = 20$ nm, $L_2 = 1.35$ m, and $P_p = 80$ mW. NL: nonlinear.

*C. Fiber length asymmetry*

The NALM principle requires a nonlinear phase difference between clockwise and counter-clockwise propagation, which is typically achieved by asymmetric passive fiber lengths $L_1$ and $L_2$ at both ends of the gain fiber inside the fiber loop. Any change in fiber length $L_2$ will not only change the NALM asymmetry but also alter the intra-cavity net dispersion and $f_{rep}$. In our study, we modified $L_2$ to change the loop asymmetry, but we kept $f_{rep}$ fixed at 54.17 MHz by simultaneously modifying the fiber length of the linear arm $L_3$ accordingly. This procedure also keeps the net cavity dispersion automatically fixed.

The simulation results are shown in Fig. 4. One can see that both the output power and the nonlinear phase difference grow with larger fiber asymmetry, whereas the double-pulsing threshold decreases. The spectral bandwidth is almost constant with fiber asymmetry, so it is not shown in Fig. 4. We did not simulate noise performance, but want to point to recent literature [30] which reports an impact of the nonlinear phase difference and therefore fiber length asymmetry on the intensity noise of the NALM oscillator. The fiber length $L_2$ was set to 1.35 m in our experimental demonstration of the laser.

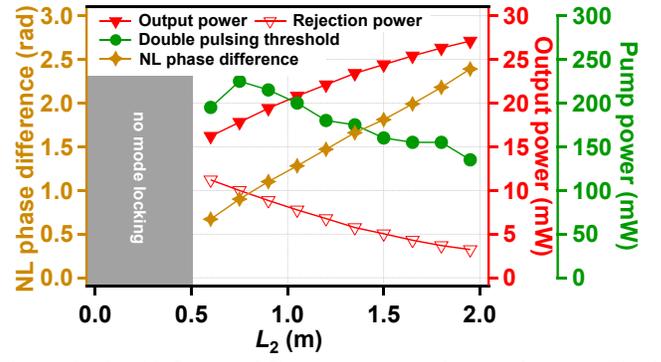

Fig. 4. Simulated influence of fiber asymmetry on laser performance. Fixed parameters: $\Phi_0 = 0.5\pi$, $r = 50\%$, $BW_{CFBG} = 20$ nm, $f_{rep} = 54.17$ MHz and $P_p = 80$ mW.

*D. CFBG bandwidth*

The CFBG acts as an intra-cavity filter which restricts the output spectral bandwidth and influences the double pulsing threshold. However, since the spectral bandwidth also strongly depends on the pump power, here we only study the output bandwidth at the highest pump power where single pulse operation could be achieved. As is shown in Fig. 5, a broader CFBG reflection bandwidth leads to an increased double-pulsing threshold, until it saturates at $BW_{CFBG} \approx 30$ nm. Simultaneously the output bandwidth increases until it saturates at $BW_{CFBG} \approx 20$ nm. In the simulation, a relatively narrow bandwidth CFBG (e.g. 6 nm < $BW_{CFBG}$ < 20 nm) could lead to an output bandwidth larger than the CFBG itself due to the nonlinear spectral broadening in fiber before the pulses reach the CFBG, and the output bandwidth could be again slightly broadened in the pigtail after the CFBG (not considered in the simulation). With respect to the larger $BW_{CFBG}$ (e.g. > 20 nm), we observed that it does not improve the output bandwidth but leads to a distorted output spectrum. The 20 nm bandwidth limit is caused by the Yb-gain bandwidth on one hand and our dispersion compensation scheme which fixed the net cavity dispersion at $-0.01$ ps$^2$ and did not compensate higher order dispersion, on the other hand. In practice, however, one should be aware that the chirp rate and bandwidth of a CFBG also influence the highest reflectance which can be achieved when writing the grating. This effect was neglected in our simulation for simplicity. While we did not simulate noise, we want to emphasize that a narrow CFBG bandwidth can act as an intra-cavity filter, which offers a restoring force to center wavelength fluctuations and leads to suppressed Gordon-Haus timing jitter [31]. In our experimental design we choose a CFBG bandwidth of 20 nm.

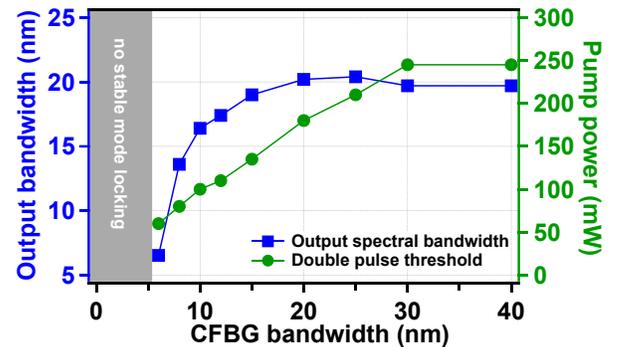

Fig. 5. Simulated influence of CFBG bandwidth on laser performance. Other parameters: $\Phi_0 = 0.5\pi$, $r = 50\%$, $L_2 = 1.35$ m, and $P_p = 5$ mW below double pulsing threshold.

*E. Other parameters*

There are still many other parameters which can influence the behavior of a mode-locked laser. An essential parameter is the net intra-cavity dispersion. We do not present the influence of intra-cavity dispersion to our NALM laser performance here, since this influence has been frequently studied in the literature both for NALM oscillators and other mode-locking mechanisms [22], [32]. In our simulations and experiments we kept the net dispersion slightly negative around zero which has been shown to be the optimum for short pulse duration and low noise [33].

Cavity loss is another crucial factor for the laser performance. We did not study this parameter, since it is governed by the maximum achievable CFBG reflectance $R_p$ for a given chirp and bandwidth. We kept this parameter fixed at $R_p = 20\%$, which is the reflectivity of our CFBGs in the experiments. Replacing the CFBG with a bulk grating pair could definitely reduce the loss, but this counteracts our goal for compactness and robustness.

As mentioned above, we fixed $f_{rep}$ for this study at 54.17 MHz to meet our application requirements. However, our design is compatible with higher $f_{rep}$, which can be achieved by simply reducing the fiber lengths. In our simulations, the highest achievable $f_{rep}$ was ~180 MHz, but the practical limit will be lower due to a required minimum component fiber pigtail length for splicing.

### III. EXPERIMENTAL SETUP AND MEASUREMENTS

*A. Laser configuration*

Fig. 6 shows the setup of the fully PM-fiber integrated oscillator we used for experiments described in the following sections, which is basically the experimental realization of the design shown in Fig. 1. We chose the parameters according to the simulation results discussed above. The loop mirror consists of a 2 × 2 PM fiber coupler with 50/50 splitting ratio, a 0.6 m length of single cladding PM Yb-doped gain fiber (YDF, Yb401-PM) which was asymmetrically placed in the loop, and a compact PM fiber integrated non-reciprocal phase shifter for starting the mode-locking operation. The pulses propagating inside the phase shifter package in clockwise and counter-clockwise directions will both double pass a λ/8 waveplate but along different axes. The different propagation axes are caused by a 45° Faraday rotator (FR), thus the clockwise and counter-clockwise pulse trains experience a π/2 phase difference. All components of the non-reciprocal phase-shifter including two collimators, one polarization beam splitter (PBS), one FR, one λ/8 waveplate and one small end mirror were pre-aligned, fixed and sealed by rigid glue in a small package as a mass-producible fiber component with 7 cm × 4 cm × 4 cm dimension, which is similar to the footprint of a fiber pigtailed terbium gallium garnet (TGG) isolator. This compact phase shifter is very robust, insensitive to environmental perturbations and has <0.5 dB insertion loss at 1030 nm.

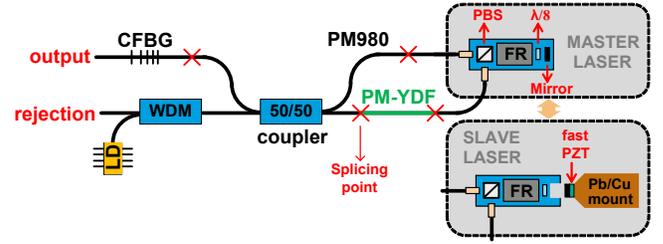

Fig. 6. Experimental scheme of the compact NALM oscillator. The red cross marks are intra-cavity fiber splicing points. CFBG: chirped fiber Bragg grating; WDM: wavelength-division-multiplexer; LD: laser diode; YDF: Yb doped fiber; PBS: polarization beam splitter; FR: Faraday rotator. PZT: piezo transducer. The slave laser, a copy of the NALM oscillator with an end mirror mounted on a PZT is used for cavity length stabilization in the later experiments.

A CFBG written in PM980 fiber was spliced into the linear arm of the cavity. We carefully designed the parameters of the CFBG to render ~20% peak reflectance centered at 1030 nm with 20 nm full-width-half-maximum (FWHM) bandwidth and to compensate the fiber dispersion such that the resulting total intra-cavity group delay dispersion ($GDD_{net}$) amounts to $-0.01$ ps$^2$, which allows the laser to mode-lock in the stretched pulse regime. The 976 nm pump light of fiber coupled LDs (1.4 W maximum power combined by two 750 mW single-mode LDs) was injected by a wavelength-division-multiplexer (WDM) through the rejection port. It should be noted that the splitting ratio of the fiber coupler at 976 nm is no longer 50/50 but around 25/75 (75 % port to the gain fiber), and the transmission of the compact phase shifter at 976 nm is only ~17 %, therefore the gain fiber was mostly pumped from one side. We prefer this pump configuration over a WDM in the NALM loop, since it minimizes the number of intra-cavity fiber splices (4 in our case) and simplifies future $f_{rep}$ scaling.

*B. Laser performance*

We could achieve mode-locking experimentally for various fiber length combinations. However, to avoid excessive values for the nonlinear phase difference (and hence to keep the double pulse threshold reasonably high), we kept the fiber asymmetry low. This design choice came with the penalty of a high pump power requirement for self-starting (~1.1 W). In the present configuration, after mode-locking is established in a multi-pulsing regime, the pump power then needs to be decreased to 150 mW to maintain stable single-pulse operation with 51 mW of average output power.

The basic output characteristics of the compact NALM oscillator are shown in Fig. 7. We obtained a smooth output optical spectrum with 19 nm FWHM bandwidth. The output pulses were positively chirped and could be compressed to 88 fs near transform-limited (TL) pulse duration using a transmission grating pair (1000 lines / mm, 10 mm grating separation). The RF spectrum of the output on a photo detector shows a carrier peak with 90 dB signal to noise ratio (SNR) at 1 kHz resolution bandwidth (RBW), and the free-running drift of $f_{rep}$ was measured to be on the sub-Hz level over a few seconds and ~40 Hz over two days. The two-day test was performed under standard laboratory conditions with a simple non-packaged and non-temperature stabilized setup when the fiber was taped to a stainless-steel breadboard and covered by an enclosure to prevent air flow, demonstrating the excellent passive stability of the design.

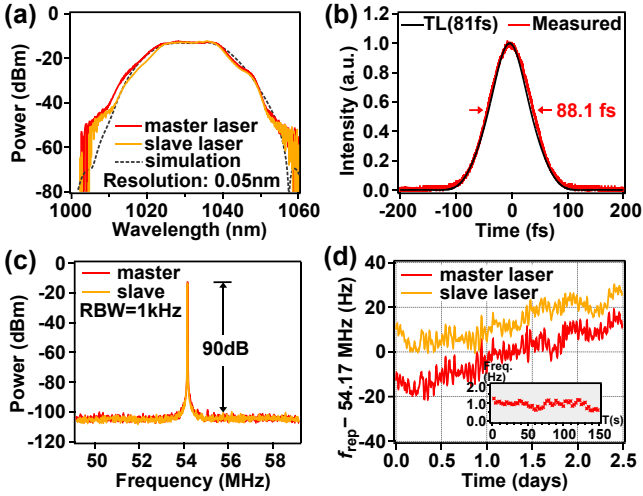

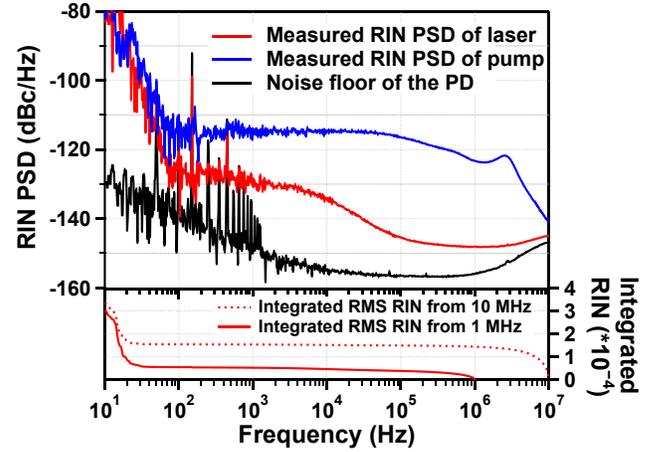

Fig. 7. Basic output performance of the compact NALM oscillators: (a) Output optical spectra and simulated spectrum; (b) compressed pulse deconvoluted from the measured auto-correlation trace by a factor of 1.41 (assuming a Gaussian pulse shape). TL: transform-limited pulse calculated from the measured optical spectrum; (c) RF spectra of $f_{rep}$; RBW: resolution bandwidth of the RF analyzer; (d) free-running drift of $f_{rep}$ over a weekend (inset shows a short-term drift).

## IV. Noise Characterizations

Low-noise operation is often an indispensable attribute of an oscillator and most crucial for both of our intended applications: seed oscillator of various timing critical laser systems at the FEL facilities and optical frequency combs. In this section, we characterize the RIN and timing jitter of our compact NALM oscillator.

### A. Relative intensity noise

The RIN was characterized by sending a part of the laser output to a Silicon photodiode (PD, BPX65 from OSRAM Opto Semiconductors). After the photocurrent was amplified and converted into a voltage signal by a trans-impedance amplifier (TIA, DHPCA-100 from FEMTO), the average voltage and baseband noise were recorded using an oscilloscope and a signal source analyzer (SSA, E5052B from Keysight Technologies), respectively. From these data we obtained the RIN power spectral density (PSD) and the corresponding integrated RIN, as shown in Fig. 8. It was determined to be $-130$ dBc/Hz at 1 kHz and $-145$ dBc/Hz at 100 kHz. The integrated RIN is $5.5 \times 10^{-5}$ (or 0.0055 %) in the integration interval [20 Hz, 1 MHz] and $1.6 \times 10^{-4}$ (or 0.016 %) in the interval [20 Hz, 10 MHz]. Extending the integration interval to 10 Hz will add another $2 \times 10^{-4}$ (0.02 %), most likely caused by the pump noise. Please also note that the RIN PSD above 3 MHz is somewhat affected by the noise floor of the PD.

Fig. 8. Measured PSD of RIN and the corresponding integration curves (from 10 MHz and from 1 MHz respectively) of the NALM oscillator output and the pump. Pump power: 150 mW; Oscillator output power: 51 mW.

### B. Timing jitter measurement

To characterize the timing jitter of the pulses, we measured the phase noise of the 24th harmonic of $f_{rep}$ at 1.3 GHz using a high-speed photo detector and the E5052B SSA. The result is shown as trace (i) in Fig. 10. However, this RF method was limited by the phase noise sensitivity of the SSA (trace (ii) in Fig.10) [34] and the photo detector [35] in the frequency range of >1 kHz. For detecting the timing jitter at higher offset frequency, a more sensitive characterization method via balanced optical cross-correlation (BOC) is required [36].

To realize this goal, we assembled another identical laser ("slave laser") which was equipped with an additional cavity length control capability. We synchronized it to the first laser ("master laser") via a BOC setup [37], depicted in Fig. 9. For cavity length tuning, we replaced the compact phase shifter with a modified version. A hole was drilled through the phase shifter package and the end mirror was mounted outside the package on a high bandwidth PZT which was rigidly glued onto a lead filled copper base [38], as shown in Fig. 6. A slow cavity tuning mechanism with larger dynamic range was constructed by coiling a part of the oscillator fiber on a piezo drum. The performance of the slave laser is also shown in Fig. 7. We observed very similar output parameters as the master laser,

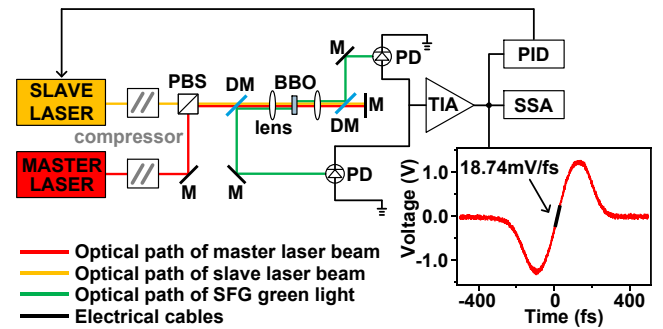

Fig. 9. BOC for timing jitter measurement of the NALM laser. The inset trace shows the error curve with a 29 Hz detuning of $f_{rep}$ between the two lasers. M: mirror; DM: dichroic mirror; BBO: type II Barium Borate crystal; PD: Silicon photodiode; TIA: trans-impedance amplifier, gain = $10^5$ V/A; PID: proportional-integral-derivative feedback control; SSA: signal source analyzer (E5052B, Keysight Technologies).

demonstrating the excellent repeatability of this compact NALM laser design.

In the BOC setup, the output pulse trains from both lasers were first compressed to sub-100 fs pulse duration and then combined in a PBS. The combined beam double-passed a type II Barium Borate (BBO) crystal for sum frequency generation (SFG). The SFG pulses of each pass were detected by a single port of a balanced photo detector (BPD). Using a TIA with $10^5$ V/A gain, we obtained the BOC error signal for synchronization and timing jitter noise characterization. The time-delay axis of the BOC error signal was calibrated by slightly detuning the $f_{rep}$ between both lasers and measuring the frequency difference $\Delta f_{rep}$ (29 Hz in our experiment) using a frequency counter while simultaneously recording the error signal on an oscilloscope. The calibrated trace and the fitted BOC sensitivity of 18.74 mV/fs are shown in the inset of Fig. 9.

For keeping the temporal overlap between the two pulse trains during the measurement, we synchronized both lasers by a low bandwidth (~25 kHz) feedback loop and characterized the BOC error signal above the servo bandwidth. Synchronization was established by a proportional-integral-derivative (PID) loop filter feedback to the cavity length of the slave laser. The measured timing jitter PSD is plotted in Fig. 10. Since the two lasers are very similar and the timing jitters of both lasers are uncorrelated, the measured timing jitter PSD was divided by a factor of 2 to obtain the contribution of a single oscillator [39]. Beyond the 25 kHz servo bandwidth, i.e. where the lasers are free-running, the jitter curve descends with a slope of $1/f^2$ and hits the noise floor at ~5 MHz. We found out that the trend is consistent with our prior RF measurement when extending the high frequency result to lower offset frequencies with the $1/f^2$ slope. The integrated root-mean-square (RMS) jitter in this [25 kHz, 5 MHz] frequency range is 0.7 fs.

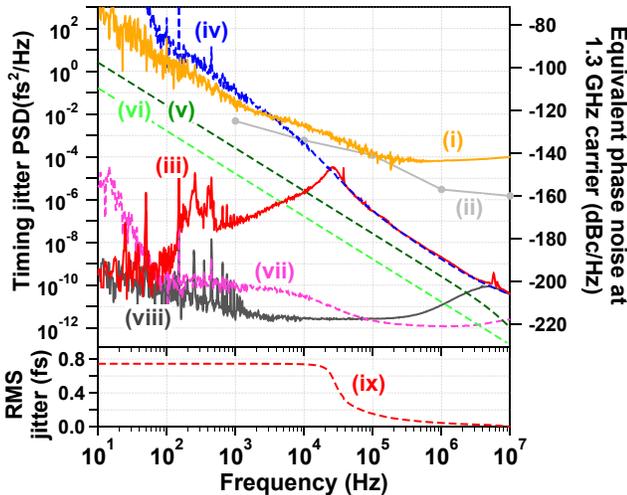

Fig. 10. (i) Measured phase noise PSD of the 24th harmonic of the $f_{rep}$ signal (1.3 GHz carrier frequency); (ii) The phase noise sensitivity of the E5052B SSA for 1 GHz carrier frequency. (iii) Measured timing jitter PSD of a single oscillator via BOC method. Note: during the BOC measurement a feedback loop with ~25 kHz servo bandwidth was active; (iv) – (vii) Calculated timing jitter PSD contributed by: (iv) RIN via self-steepening (assuming the round-trip nonlinear phase shift is 17.5 rad calculated by simulation), (v) Gordon-Haus jitter, (vi) direct influence of ASE noise (assuming the excess noise factor is 10) and (vii) RIN via Kramers-Kronig relations, respectively; (viii) Noise floor of the BPD; (ix) Integrated RMS jitter of curve (iii).

To visualize the dominating contributions and limitations of the timing jitter, we also plotted the calculated the jitter PSD traces [31], [40] from several different contributions in Fig. 10. It can be seen that the timing jitter of our oscillator is currently limited by the RIN-coupled self-steepening effect. Even at our very low RIN level this effect is still dominating, since the self-steepening induced timing jitter is also proportional to the square of the total nonlinear phase shift in one round-trip, which for our case is as high as 17.5 rad according to the simulation. Second highest contribution is the Gordon-Haus jitter, which is calculated to be ~10 dB lower than the self-steepening induced jitter. We attribute this to the residual non-zero intra-cavity dispersion and the CFBG bandwidth, which is broader than the laser spectral bandwidth.

Further suppression of the timing jitter of this oscillator could possibly be achieved by sacrificing a part of the output power to avoid very high nonlinear phase shifts and by utilizing a better designed CFBG with more precise dispersion compensation and narrower bandwidth to suppress the Gordon-Haus jitter. Additionally, an electronic RIN eater circuit could also be beneficial [41], [42]. With all measures mentioned above in place, the timing jitter of our oscillator design will be limited by jitter directly induced by amplified spontaneous emission (ASE) noise. This contribution is high compared to those high cavity-Q solid-state oscillators due to the intrinsically high output coupling ratio (80 %) of the CFBG, which is nevertheless an inevitable trade-off for the robust and compact design.

## V. OPTICAL FREQUENCY COMB

OFCs are powerful tools in many advanced research fields. In this section, we show that our NALM oscillator can be fully phase stabilized and used for OFC applications.

It is well known that an OFC has two degrees of freedom, the carrier-envelope offset frequency $f_{ceo}$ and the repetition frequency $f_{rep}$, both of which need to be stabilized. In our experiment, a conventional $f$-to-$2f$ interferometer (not shown) was built to detect the $f_{ceo}$ signal, which was then phase-locked to a stable RF synthesizer by a phase-locked loop (PLL) feeding back to the driving current of the pump LD. Alternatively to directly stabilizing $f_{rep}$, a heterodyne beat note ($f_{beat}$) was obtained by beating the NALM laser with a narrow-linewidth continuous-wave laser (Koheras ADJUSIK Y-10 from NKT Photonics, <20 kHz specified linewidth) at 1030 nm (~ 291 THz). Another PLL was employed to phase lock the $f_{beat}$ to a stable RF reference by driving the fast PZT (currently only available in the slave laser).

Figure 11 shows the in-loop locking results of our fully phase stabilized frequency comb. Both signals (insets of Fig. 11) show distinct coherent peaks on the RF spectra, indicating a tight phase lock. The phase noise PSDs of both signals were integrated, resulting in 0.12 rad RMS and 0.4 rad RMS phase error for $f_{ceo}$ and $f_{beat}$, respectively, in the integration interval [10 Hz, 5 MHz]. With the estimated feedback bandwidth of 22 kHz and the fact that the integrated $f_{beat}$ phase noise in the interval [22 kHz, 5 MHz] is only ~0.3 rad, we also conclude that 22 kHz will be an upper limit for the free-running comb linewidth, since this value is well below the 1 rad value which can be used to estimate the linewidth [43].

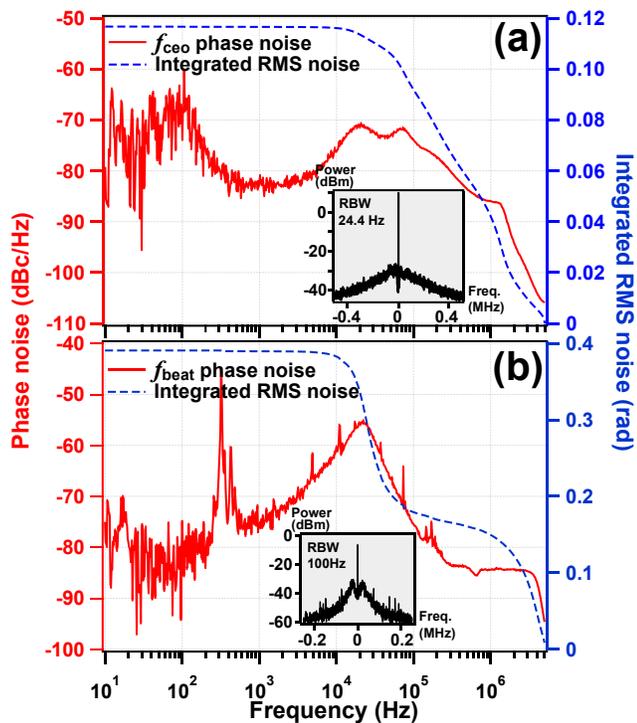

Fig. 11. Phase noise PSD and integrated RMS phase error of (a) $f_{ceo}$ signal and (b) $f_{beat}$ after locking (insets show the corresponding RF spectra). RBW: resolution bandwidth of the RF analyzer.

## VI. Discussion and outlook

While our $f_{rep}$ fixed "master laser" oscillator is alignment-free, the cavity length feedback used in the "slave laser" still contains free-space parts which require manual alignment. However, the mirror-PZT-mount assembly can easily be integrated into the compact phase shifter, resulting in an alignment-free frequency comb oscillator with cavity length stabilization and $f_{ceo}$ control options. We have planned this upgrade in the near future.

As an alternative method for high speed cavity length feedback control, an electro-optic modulator (EOM) can be implemented into the laser cavity [44], allowing feedback bandwidths in the MHz range. Although utilizing the fiber integrated waveguide-type EOMs for phase locking was to the best of our knowledge only reported in the Er:fiber laser OFCs, we have experimentally confirmed stable mode-locking operation of our all-fiber Yb NALM laser when a waveguide-type EOM with PM980 fiber pigtails was spliced into the cavity. We were also able to modulate this waveguide-type EOM and generated clean sidebands on the RF spectrum of $f_{rep}$, which indicate the possibility of phase locking the $f_{rep}$ (or $f_{beat}$) of the Yb:fiber NALM laser by the waveguide-type EOM.

In addition to the low timing jitter and excellent frequency comb performance, our oscillator design combines several technical advantages simultaneously: alignment free operation, insensitivity to environmental fluctuations, no observable component degradation over months, robustness, compactness, very good reproducibility and easy assembly. In fact, assembling the laser is as simple as cutting the component fiber pigtail leads to the correct length and executing four intra-cavity standard PM fiber splices plus two more for the WDM and the pump LD.

To test the robustness, we taped the fiber oscillator on a hand-size aluminum base plate and initiated a free fall of the running oscillator from ~15 cm height. No change was observed on the optical spectrum, RF spectrum and time domain pulse train. Also, the running laser was immune to shock impact as tested using a hammer (see supplementary material for the video). While those tests were not well defined and somewhat arbitrary, they nevertheless demonstrate the excellent environmental stability of this laser.

## VII. Conclusion

We have demonstrated a compact and alignment-free femtosecond Yb:fiber oscillator mode-locked by a NALM. The influence of several cavity design parameters (phase bias, coupling ratio, fiber asymmetry and CFBG bandwidth etc.) on laser performance has been studied in numerical simulations. Key laser performance aspects, such as output power, bandwidth, difficulty of initiating mode-locking, and double pulsing have been analyzed. Our numerical simulations provided a guideline for the parameter choice in the experiments.

In a second step, we realized the oscillator experimentally and demonstrated 51 mW average output power at 54.17 MHz repetition rate and 19 nm FWHM optical bandwidth centered at 1030 nm. The pulses could be compressed to 88 fs in a clean shape by a simple grating pair compressor. We characterized the RIN and timing jitter to be $5.5 \times 10^{-5}$ [20 Hz – 1 MHz] and 0.7 fs RMS [25 kHz – 5 MHz], respectively. We finally demonstrated that the oscillator can be tightly phase-stabilized in $f_{ceo}$ and to a 291 THz optical reference for frequency comb applications with residual RMS phase errors of 0.12 rad ($f_{ceo}$) and 0.4 rad ($f_{beat}$), respectively in the integration interval from 10 Hz to 10 MHz.

The laser is constructed in a rugged design, is environmentally stable and can be easily assembled with good reproducibility. We are convinced that such a reliable Yb:fiber femtosecond oscillator meets all requirements for our intended applications in FEL science and frequency comb spectroscopy, and furthermore can be an ideal femtosecond seed source for high power Yb-laser systems and OFCs serving other applications.